\documentclass[a4paper,final]{My_style}
\usepackage{amsmath}
\usepackage[english]{babel}
\usepackage{cite}
\usepackage[dvips]{color}
\usepackage{dsfont}
\usepackage{fancyhdr}
\usepackage{theorem}
\usepackage{txfonts}

\newcommand{\pt}[1]{\left( #1 \right)}
\newcommand{\pq}[1]{\left[ #1 \right]}
\newcommand{\pg}[1]{\left\{ #1 \right\}}
\newcommand{\bs}[1]{\boldsymbol{#1}}
\newcommand{\coh}[2]{\ensuremath{|C_{#1}(#2) \rangle}}
\newcommand{\lcoh}[2]{\ensuremath{\langle C_{#1}(#2)|}}
\newcommand{\floor}[1]{\left\lfloor #1 \right\rfloor}
\renewcommand{\c}[1]{\mathcal{#1}}
\newcommand{\g}[1]{\mathfrak{#1}}
\newcommand{\co}[1]{\textsf{#1}}
\DeclareMathOperator*{\tr}{Tr}

{ \theorembodyfont{\rmfamily}
\newtheorem{remark}{Remark}[section]}
\newtheorem{definition}{Definition}[section]
\newtheorem{proposition}{Proposition}[section]
\newtheorem{condition}{Condition}[section]
\newtheorem{theorem}{Theorem}[section]

\begin{document}

\pagestyle{fancy}
\renewcommand{\headrulewidth}[0]{0pt}

 \lhead[\fancyplain{}{ \thepage}]
       {\fancyplain{}{}}
 \chead[\fancyplain{}{ {\it V.~Cappellini}}]
       {\fancyplain{}{ {\it A Survey on the Classical Limit of Quantum Dynamical Entropies}}}
 \rhead[\fancyplain{}{}]
       {\fancyplain{}{ \thepage}}

 \lfoot{}\cfoot{}\rfoot{}

\title{\vspace{-2cm}\bf\huge{}A Survey on the Classical Limit\\
{\bf\huge{}of Quantum Dynamical Entropies\rule[0pt]{0pt}{24pt}}%
\thanks{Proceedings of the  $3$rd Workshop on
Quantum Chaos and Localisation Phenomena,\\
Warsaw, Poland, May 25--27, 2007}
 }
\author{\rule[0pt]{0pt}{28pt}\Large V.~Cappellini
\address{%
\normalsize\rule[0pt]{0pt}{20pt}
``Mark Kac'' Complex Systems Research Centre, Uniwersytet Jagiello{\'n}ski\\[0.3ex]
ul. Reymonta 4, 30--059 Krak\'{o}w, Poland\\[0.9ex]
and\\[0.9ex]
Centrum Fizyki Teoretycznej, Polska Akademia Nauk\\[0.3ex]
Al. Lotnik{\'o}w 32/44, 02--668 Warszawa, Poland\\[-6ex]
\ }}
 \maketitle
 \begin{abstract}
We analyze the behavior of quantum dynamical entropies production
from sequences of quantum approximants  approaching their (chaotic)
classical limit. The model of the quantized hyperbolic automorphisms
of the 2--torus is examined in detail and a semi--classical analysis
is performed on it using coherent states, fulfilling an appropriate
dynamical localization property. Correspondence between quantum
dynamical entropies and the Kolmogorov--Sinai invariant is found
only over time scales that are logarithmic in the quantization
parameter.
\end{abstract}
\PACS{05.45.-a, 05.45.Mt, 05.45.Ac, 03.65.Fd}
\tableofcontents
\section{Introduction}
\label{s1} \ \\[-6ex]

 \indent The notion of classical chaos is associated with
motion on a compact phase--space with high sensitivity to initial
conditions: trajectories diverge exponentially fast and nevertheless
remain confined to bounded
regions~\cite{Dev89:1,Wig90:1,Kat99:1,Sch95:1,Gia89:1,Cas95:1,Zas85:1}.

In an opposite way, quantization on compacts yields discrete energy
spectra, which in term entail quasi--periodic
time--evolution~\cite{For92:1}.

Nevertheless, nature is fundamentally quantal and, according to the
correspondence principle, classical behavior must emerge in the
limit $\hbar\to0$.

{ Also, classical and quantum mechanics are expected to overlap over
times expected to scale as $\hbar^{-\alpha}$ for some
$\alpha>0$~\cite{Zas85:1}, the so--called semi--classical regime.
Actually, it turns out that this is true only for regular classical
limits whereas, for chaotic ones, classical and quantum mechanics
agree over times which scale as
$-\log\hbar$~\cite{Gia89:1,Cas95:1,Zas85:1}, and footprints of the
exponential separation of classical trajecto\-ries are found even on
finite dimensional quantization provide that the time does not
exceed such a logarithmic upper bound~\cite{Cas95:1,Zas91:1}.} Both
time scales diverge when $\hbar\to0$, but the shortness of the
latter means that classical mechanics has to be replaced by quantum
mechanics much sooner for quantum systems with chaotic classical
behavior. The logarithmic breaking time $-\!\log{\hbar}$ has been
considered by some as a violation of the correspondence
principle~\cite{For91:1,For92:2} and by others, see~\cite{Cas95:1}
and Chirikov in~\cite{Gia89:1}, as the evidence that time and
classical limits do not commute.

The analytic studies of logarithmic time scales have been mainly
performed  by means of semi--classical tools, essentially by
focusing, via coherent state techniques,  on the phase space
localization  of specific time evolving quantum observables. In the
following, we shall show how they emerge in the context of quantum
dynamical entropies.

As a particular example, we shall concentrate on finite dimensional
quantizations of continuous hyperbolic automorphisms of the 2--torus
$\mathds{T}^2\coloneqq\mathds{R}^2/\mathds{Z}^2$ (the unit square
with opposite sides identified), which are prototypes of chaotic
behavior; indeed, their trajectories separate exponentially fast
with a Lyapunov exponent $\log\lambda_{+}>0$~\cite{Arn68:1,Wal82:1}.
 If $\delta$ is an initial error along a trajectory, and
$\delta_n\simeq\delta\lambda_{+}^n$ its classical spreading after
$n$ steps of the (time--stroboscopic) dynamics, then boundness of
the motion imposes $\delta_n\leqslant 1$, where 1 is the diameter of
the 2--torus $\mathds{T}^2$. This explain why the limit $\delta\to0$
has necessarily to be performed before the time--limit, and the
Lyapunov exponent can be computed as
\begin{equation}
\log\lambda_{+}=\lim_{n\to\infty}\;\frac{1}{n}\; \lim_{\delta\to
0}\;\log\pt{{\delta_n\over\delta}}\quad.\label{lyap_exp}
\end{equation}
Standard quantization, \`a la Berry, of hyperbolic automorphisms on
$\mathds{T}^2$~\cite{Ber79:1,Deg93:1} yields Hilbert spaces of a
finite dimension~$N$, this latter variable playing the role of the
semi--classical parameter and setting to $1/N$ the minimal size of
the phase--space grain cells. Imposing the latter bound,
$\min\pg{\delta}\geqslant 1/N$, its evident how the conflict between
the two limits, emerging once $\delta_n\simeq 1$, can be transferred
in the time--step $n$ as $n\simeq \log N/\log\lambda_{+}$. In this
sense, rather than a violation of the correspondence principle, the
logarithmic breaking--time indicates the typical scaling for a joint
time--classical limit suited to classically chaotic quantum systems.

The \co{K}olmogorov--\co{S}inai dynamical entropy~\cite{Kat99:1}
(\co{KS}--entropy, for short) is defined by assigning measures to
bunches of trajectories and computing the Shannon--entropy per
time--step of the ensemble of bunches in the limit of infinitely
many time--steps: The more chaotic the time--evolution, the more the
possibile bunches and the larger their entropy. The production of
different bunches of trajectories issuing from the same bunch is
typical of high sensitivity to initial conditions and this is indeed
the mechanism at the basis of the theorem of Ruelle and
Pesin~\cite{Man87:1}, linking \co{KS}--entropy of a smooth,
classical dynamical systems, to the sum of its positive Lyapunov
exponents.

In the quantum realm, there are different candidates for
non--commutative extensions of the
\co{KS}--invariant~\cite{Con87:1,Ali94:1,Voi92:1,Acc97:1,Slo94:1}:
in this paper we shall focus on one of them, called
\co{ALF}--entropy~\cite{Ali94:1}, and we shall study its
semi--classical limit.

The \co{ALF}--entropy is based on the algebraic properties of
dynamical systems, that is on the fact that they are describable by
suitable algebras of observables, their time evolution by linear
maps on these algebras, and their states by expectations over them.

We show that, while being bounded by $\log N$, nevertheless over
numbers of time steps $1\ll n<\log N$, the entropy content per
letter, or entropy production, is $\log\lambda_{+}$\,. It thus
follows that the joint limit $n,N\to+\infty$, with $n\propto\log N$,
yields the \co{K}olmogorov--\co{S}inai  entropy.{} This confirms the
numerical results in~\cite{Ali96:2} and~\cite{Pog07:1}, where the
dynamical entropy~\cite{Ali94:1} is applied to the study of  the
quantum kicked top, respectively to quantum cat maps.

In this approach, the presence of logarithmic time scales indicates
the  typical scaling for a joint time/classical limit suited to
preserve positive  entropy production in quantized classically
chaotic quantum systems.

The paper is organized as follows: Section~\ref{s2} contains a brief
review of the algebraic approach to classical and dynamical systems,
while Section~\ref{s3} introduces some basic semi--classical tools.
Sections~\ref{s4} and~\ref{s5} deal with the quantization of
hyperbolic maps on finite  dimensional Hilbert spaces and the
relation between classical and time limits. Section~\ref{s6} gives
an overview of various models of quantum dynamical entropies present
in the literature and particularly focus on the one proposed by
Alicki and Fannes~\cite{Ali01:1,Ali94:1} (\co{ALF}--entropy, where
\co{L} stands for Lindblad). Finally, in Section~\ref{s7}, the
semi--classical behavior of quantum dynamical entropies is studied
and the emergence of a typical logarithmic time scale is showed.

\section{Dynamical systems: algebraic setting}
\label{s2}

 Usually, continuous classical motion is described by means of
a measure space ${\cal X}$, the phase--space, endowed with the Borel
$\sigma$--algebra and a normalized measure $\mu$, $\mu({\cal X})=1$.
The ``volumes''
\begin{equation*}
\mu(E)=\int_E\mu(\mathrm{d}\bs{x})
\end{equation*}
of measurable subsets $E\subseteq{\cal X}$ represent the
probabilities that phase--points $\bs{x}\in{\cal X}$ belong to them.
By specifying the statistical properties of the system, the measure
$\mu$ defines a ``state'' of it. In such a scheme, a reversible
discrete time dynamics amounts to an invertible measurable map $T$
onto ${\cal X}$ such that $\mu\circ T=\mu$, and to its iterates
$\{T^k \mid k\in\mathds{Z}\}$: $T$--invariance of the measure $\mu$
ensure that the state defined by $\mu$ can be taken as an
equilibrium state with respect to the given dynamics.
Phase--trajectories passing through $\bs{x}\in{\cal X}$ at time $0$
are then sequences $\{T^k \bs{x}\mid
k\in\mathds{Z}\}$~\cite{Kat99:1}.

\noindent Classical dynamical systems are thus conveniently
described by triplets $({\cal X},T,\mu)$. In the present work we
shall focus upon the following:
\begin{itemize}
\item
 $\c X$ -- a compact metric space:\\ the $2$--dimensional
 torus $\mathds{T}^2={\mathds{R}}^2/{\mathds{Z}}^2 =\{(x_1,x_2)\in\mathds{R}^2\
 \pmod{1} \}$;
\item
 $T$ -- invertible measurable transformations from $\c X$ to itself
 such that $T^{-1}$ are also measurable;
\item
 $\mu$ -- the Lebesgue measure
 $\mu(\mathrm{d}\bs{x})=\mathrm{d} x_1\,\mathrm{d} x_2$ on $\mathds{T}^2$.
\end{itemize}

In this paper, we consider a general scheme for quantizing and
dequantizing, i.e.\ for taking the classical limit
(see~\cite{Wer95:1}). Within this framework, we focus on the
semi--classical limit of quantum dynamical entropies of finite
dimensional quantizations of the celebrated Arnold's cat map and of
generic { maps belonging to the so--called \emph{unimodular group}
on the 2--torus: in the following we simply denote such a family of
maps \emph{cat maps family}. The last denomination is perfectly
legitimate, in fact the acronym \co{CAT} stands for \co{C}ontinuous
\co{A}utomorphism of the \co{T}orus.}

In order to make the quantization procedure more explicit, it proves
useful to follow an algebraic approach and replace $(\c X,T,\mu)$
with $(\g M_\mu,\Theta,\omega_\mu)$ where
\begin{itemize}
\item
 $\g M_\mu$ -- is the von~Neumann algebra $\g L_ \mu^\infty(\c X)$ of
 (equivalence classes of) essentially bounded $\mu$--measurable
 functions on $\c X$, equipped with the so--called essential supremum
 norm $\|\cdot\|_\infty$~\cite{Rud87:1};
\item
 $\{\Theta^k \mid k\in\mathds{Z}\}$ -- is the discrete group of automorphisms
 of $\g M_\mu$ which implements the dynamics: $\Theta(f) \coloneqq f \circ
 T^{-1}$. The invariance of the reference measure reads now
 $\omega_\mu \circ \Theta = \omega_\mu$\;;
\item
 $\omega_\mu$ -- is the state on $\g M_\mu$ defined by the reference
 measure $\mu$
\begin{equation*}
\omega_\mu:{\g M_\mu}\ni
f\longmapsto\omega_\mu(f)\coloneqq\int_{{\cal X}} \mu\pt{\mathrm{d}
\bs{x}}\ f(x)\in \mathds{R}^{+}\quad .\nonumber
\end{equation*}

\end{itemize}

Quantum dynamical systems are described in a completely similar way
by a triple $(\g M,\Theta,\omega)$, the critical difference being
that the algebra of observables $\g M$ is no longer Abelian:
\begin{itemize}
\item
 $\g M$ -- is a von~Neumann algebra of operators, the observables, acting on
 a Hilbert space $\g H$\;;
\item
 $\Theta$ -- is an automorphism of $\g M$\;;
\item
 $\omega$ -- is an invariant normal state on $\g M$: $\omega\circ\Theta=\omega$\;.
\end{itemize}

Quantizing essentially corresponds to suitably mapping the
commutative, classical triple $(\g M_\mu,\Theta,\omega_\mu)$ to a
non--commutative, quantum triple $(\g M,\Theta,\omega)$.

\section{Classical limit: coherent states}
\label{s3}

 Performing the classical limit or a semi--classical analysis
consists in studying how a family of algebraic triples $(\g
M,\Theta,\omega)$, depending on a quantization $\hbar$--like
parameter, is mapped onto $(\g M_\mu,\Theta,\omega_\mu)$ when the
parameter goes to zero. The most successful semi--classical tools
are based on the use of coherent states (\co{CS} for short).

For our purposes, we shall use a large integer $N$ as a quantization
parameter, i.e.\ we use $1/N$ as the $\hbar$--like parameter. In
fact, we shall consider cases where $\g M$ is the algebra $\c M_N$
of $N$--dimensional square matrices acting on $\mathds{C}^N$, the
quantum reference state is the normalized trace $\frac{1}{N} \tr\ $
on $\c M_N$, denoted by $\tau_N$, and the dynamics is given in terms
of a unitary operator $U_T$ on $\mathds{C}^N$ in the standard way:
$\Theta_N(X) \coloneqq U_T^*X\,U_T^{\phantom{*}}$.

In full generality, coherent states will be identified as follows.

\begin{definition}
\label{coh}
 A family $\{\vert C_N(\bs{x})\rangle \mid \bs{x}\in\c X\}\in\g H$ of vectors,
 indexed by points
 $\bs{x}\in\c X$, constitutes a set of coherent states if it satisfies the
 following requirements
 \begin{enumerate}
 \item
  \co{\upshape Measurability}: $\bs{x} \mapsto \vert C_N(\bs{x})\rangle$ is measurable on $\c X$;
 \item
  \co{\upshape Normalization}: $\|C_N(\bs{x})\|^2 = 1$, $\bs{x}\in\c X$;
 \item
  \co{\upshape Overcompleteness}: $N \int_{\c X}\mu(\mathrm{d} \bs{x})\, \coh{N}{\bs{x}}
  \lcoh{N}{\bs{x}} = \mathds{1}_N$;
 \item
  \co{\upshape Localization}: given $\varepsilon>0$ and $d_0>0$, there exists
  $N_0(\epsilon,d_0)$ such that for $N\ge N_0$ and $d_{\cal X}(\bs{x},\bs{y})\ge d_0$ one has
  \begin{equation*}
   N |\langle{} C_N(\bs{x}), C_N(\bs{y}) \rangle|^2 \le \varepsilon.
  \end{equation*}
 \end{enumerate}
\end{definition}
\noindent The symbol $d_{\cal X}(\bs{x},\bs{y})$ used in the
\co{localization} property stands for the length of the shorter
segment connecting the two points on ${\cal X}$. Of course the
latter quantity does depend on the topological properties of ${\cal
X}$. In particular, for the 2--torus,
\begin{equation}
d_{\mathds{T}^2}\pt{\bs{x},\bs{y}}  \coloneqq
\min_{\bs{n}\in{\mathds{Z}}^2}
\big\|\:\bs{x}-\bs{y}+\bs{n}\;\big\|_{{\mathds{R}}^2}
\label{Gnbar_m2}\quad.
\end{equation}

The \textsf{overcompleteness} condition may be written in dual form
as
\begin{equation*}
 N \int_{\c X}\mu(\mathrm{d} \bs{x})\, \langle{}C_N(\bs{x}), X\, C_N(\bs{x})\rangle = \tr X, \quad X\in\c
 M_N.
\end{equation*}
Indeed,
\begin{equation*}
 N \int_{\c X}\mu(\mathrm{d} \bs{x})\, \langle{}C_N(\bs{x}), X\, C_N(\bs{x})\rangle = N \tr \pt{\int_{\c
 X}\mu(\mathrm{d} \bs{x})\, \coh{N}{\bs{x}} \lcoh{N}{\bs{x}}\, X} = \tr X\quad.
\end{equation*}

\subsection{Anti--Wick Quantization}

In order to study the classical limit and, more generally, the
semi--classical behavior of $(\c M_N ,\Theta_N,\tau_N)$ when
$N\to\infty$, we introduce two linear maps. The first,
$\gamma_{N\infty}$, (anti--Wick quantization) associates $N\times N$
matrices of $\c M_N$ to functions in $\g M_\mu=\g L_ \mu^\infty(\c
X)$; the second one, $\gamma_{\infty N}$, maps $N\times N$ matrices
to functions in $\g L_ \mu^\infty(\c X)$.

\begin{definition}
\label{qWick}
 Given a family $\{\;\vert C_N(\bs{x})\rangle \mid \bs{x}\in\c X\;\}$ of \co{CS} in
 $\mathds{C}^N$, the anti--Wick quantization scheme will be described by a
 (completely) positive unital map $\gamma_{N\infty}: \g M_\mu\to\c
 M_N$
 \begin{equation*}
 { \g M_\mu\ni} f \mapsto
 N \int_{\c X}\mu(\mathrm{d} \bs{x})\, f(\bs{x})\,
  \coh{N}{\bs{x}} \lcoh{N}{\bs{x}}\eqqcolon\gamma_{N\infty}(f)\in \c M_N\quad .
 \end{equation*}
 The corresponding dequantizing map $\gamma_{\infty N}: \c M_N\to\g M_\mu$
 will correspond to the (completely) positive unital map
 \begin{equation*}
  \c M_N \ni X \mapsto
  \langle{}C_N(\bs{x}), X\,C_N(\bs{x})\rangle\eqqcolon\gamma_{\infty N}(X)(\bs{x}) \in \g M_\mu
  \quad .
 \end{equation*}
\end{definition}

Both maps are identity preserving because of the conditions imposed
on the \co{CS}--family of and are also completely positive, since
the domain of $\gamma_{N\infty}$ is a commutative algebra as well as
the range of $\gamma_{\infty N}$\;. The following two equivalent
properties are less trivial:

\begin{proposition}
\label{prop1}
 For all $f\in\g M_\mu$
 \begin{equation*}
  \lim_{N\to\infty} \gamma_{\infty N} \circ \gamma_{N\infty}(f) =
  f\quad \mu\,\text{--}\,{\mathrm{a.\ e.}}
 \end{equation*}
\end{proposition}
\medskip

\begin{proposition}
\label{prop2}
 For all $f,g\in\g M_\mu$
 \begin{equation*}
  \lim_{N\to\infty} \tau_N \bigl( \gamma_{N\infty}(f)^*
  \gamma_{N\infty}(g) \bigr) = \omega_\mu(\overline f g) = \int_{\c X}
  \mu(\mathrm{d} \bs{x})\, \overline{f(\bs{x})}g(\bs{x}).
 \end{equation*}
\end{proposition}
\medskip

\noindent The previous two propositions, proved
in~\cite{Ben03:1,Cap04:1}, can be taken as requests on any
well--defined quantization/dequantization scheme for observables. In
the sequel, we shall need the notion of quantum dynamical systems
$(\c M_N,\Theta_N,\tau_N)$ tending to the classical limit $(\g
M_\mu,\Theta,\omega_\mu)$. We then not only need convergence of
observables but also of the dynamics. This aspect will be considered
in Section~\ref{s5}.

\medskip

\section{Classical and quantum cat maps}
\label{s4}

In this section, we collect the basic material needed to describe
both classical and quantum cat maps and we introduce a specific set
of \co{CS} that will enable us to perform the semi--classical
analysis over such dynamical systems.

\subsection{Finite dimensional quantizations}

We first introduce cat maps in the spirit of the algebraic
formulation introduced in the previous sections.

\begin{definition}
 \label{ccat-map}
 Hyperbolic continuous automorphisms of the torus are generically
 represented by triples $(\g M_\mu,\Theta,\omega_\mu)$, where
 \begin{itemize}
 \item
  $\g M_\mu$ is the algebra of
  essentially bounded functions on the two dimensional torus
  $\mathds{T}^2\coloneqq\mathds{R}^2/\mathds{Z}^2 = \bigl\{(x_1,x_2)\in\mathds{R}^2\ \pmod{1} \bigr\}$, equipped
  with the Lebesgue measure $\mu(\mathrm{d}\bs{x}) \coloneqq \mathrm{d}\bs{x}$\,;
 \item
  $\{\Theta^k \mid k\in\mathds{Z}\}$ is the family of automorphisms
  (discrete time evolution) given by
  $\g M_\mu\ni f \mapsto (\Theta^kf)(\bs{x}) \coloneqq f(A^{-k}\bs{x}\pmod{1})$,
  where $A=\pt{\begin{smallmatrix} a & b \\ c & d\end{smallmatrix}}$ has integer entries
  such that $ad-bc=1$, $|a+d|>2$ and maps $\mathds{T}^2$ onto itself\,;
 \item
  $\omega_\mu$ is the expectation obtained by integration
  with respect to the Lebesgue measure:
  $\g M_\mu\ni f\mapsto\omega_\mu(f)\coloneqq\int_{\mathds{T}^2} \mathrm{d}\bs{x}\, f(\bs{x})$, that is
  left invariant by $\Theta$\,.
 \end{itemize}
\end{definition}

\noindent Denoting with $t\coloneqq\tr\pt{A}/2$ the semi--trace of
$A$, $\left|t\right|>1$, the two irrational eigenvalues of $A$ can
be written as $1<\lambda_{+}^{\phantom{1}}\coloneqq t+\sqrt{t^2-1}$
and $1> \lambda_{-}^{\phantom{1}}\coloneqq
t-\sqrt{t^2-1}=\lambda_{+}^{-1}$. Distances are stretched along the
direction of the eigenvector $|\bs{e}_+\rangle$,
$A\,|\bs{e}_+\rangle=\lambda_+|\bs{e}_+\rangle$, contracted along
that of $|\bs{e}_-\rangle$, $A\,|\bs{e}_-\rangle=\lambda_-
|\bs{e}_-\rangle$ and all periodic points are
hyperbolic~\cite{Per87:1}. Once the folding condition is added, the
hyperbolic automorphisms of the torus become prototypes of classical
chaos, with positive Lyapunov exponent $\log\lambda_{+}$.

One can quantize the associated algebraic triple $(\g
M_\mu,\Theta,\omega_\mu)$ on either infinite~\cite{Ben91:2} or
finite dimensional Hilbert spaces~\cite{Ber79:1,Deg93:1,Deb98:1}. In
the following, we shall focus on the latter.

Given an integer $N$, we consider an orthonormal basis $\vert
j\rangle$ of $\mathds{C}^N$, where the index $j$ runs through the
residual class modulo $N$, here and in the following denoted by
$(\mathds{Z}/ N \mathds{Z})$, namely $|j+N\rangle \equiv |j\rangle,
\ j \in \mathds{Z}$. By using this basis we define two unitary
matrices $U_N$ and $V_N$, representing position and momentum
\emph{shift operators}, as follows:
\begin{equation}
 U_N\vert j\rangle \coloneqq \exp{\pt{\frac{2\pi i}{N}u}}\vert j+1\rangle,
 \qquad\text{and}\qquad
 V_N\vert j\rangle \coloneqq \exp{\pt{\frac{2\pi i}{N}(v-j)}} \vert j\rangle.
\label{UV}
\end{equation}
In the last equation, we explicitly indicated the dependence on two
arbitrary phases $(u,v)\in [0,1)$ labeling the representation and
fulfilling
\begin{equation}
  U_N^N = \mathrm{e}^{2i\pi u}\, \mathds{1}_N,\quad
  V_N^N = \mathrm{e}^{2i\pi v}\, \mathds{1}_N.
\label{FoldUV}
\end{equation}
It turns out that
\begin{equation}
 U_NV_N=\exp{\pt{\frac{2i\pi}{N}}}\, V_NU_N.
\label{algUV}
\end{equation}
Introducing Weyl operators labeled by $\bs{n} =
(n_1,n_2)\in\mathds{Z}^2$
\begin{equation}
  W_N(\bs{n})\coloneqq \exp{\pt{\frac{i\pi}{N}n_1n_2}}\, V_N^{n_2}U_N^{n_1} =
  W_N(-\bs{n})^*
\label{Weyl1}
\end{equation}
it follows that
\begin{subequations}\label{Weyl222-333}
\begin{align}
 W_N(N\bs{n})
 &= \mathrm{e}^{i\pi(N n_1n_2+ 2n_1 u+ 2n_2 v)}
\label{Weyl2} \\
 W_N(\bs{n})W_N(\bs{m})
 &= \exp{\pt{\frac{i\pi}{N}\sigma(\bs{n},\bs{m})}}\, W_N(\bs{n}+\bs{m}),
\label{Weyl3}
\end{align}
\end{subequations}
 where $\sigma(\bs{n},\bs{m}) \coloneqq n_1m_2-n_2m_1$
is the so--called symplectic form.

\begin{definition}
\label{qcat-map-def}
 Quantized cat maps will be identified with triples
 $(\c M_N ,\Theta_N,\tau_N)$ where
 \begin{itemize}
 \item
  $\c M_N $ is the full $N\times N$ matrix algebra over
$\mathds{C}$ generated by the (discrete) group of Weyl operators
$\{W_N(\bs{n})\mid \bs{n}\in{\mathds{Z}}^2\}$\,. In the following,
such a group will be denoted by \emph{Weyl group\;;}
 \item
  $\Theta_N: \c M_N\mapsto \c M_N$ is
  the automorphism such that
  \begin{equation}
   W_N(\bs{p}) \mapsto\Theta_N(W_N(\bs{p}))\coloneqq W_N(A\bs{p})\quad,\quad
  \bs{p}\,\in(\mathds{Z}/ N \mathds{Z})^2\quad.
  \label{dyncat}
  \end{equation}
 \end{itemize}
\end{definition}
In the definition of above, we have omitted reference to the
parameters $u,v$ in~(\ref{UV}): they must be chosen such that
 \begin{equation}
  \begin{pmatrix} a & c \\ b & d \end{pmatrix}
  \begin{pmatrix} u \\ v \end{pmatrix} =
  \begin{pmatrix} u \\ v \end{pmatrix} +
  \frac{N}{2} \begin{pmatrix} ac \\ bd \end{pmatrix}\pmod{1}\quad.
 \label{uv}
 \end{equation}
Then, the folding condition~\eqref{FoldUV} is compatible with the
time evolution~\cite{Deg93:1}. The reason for~\eqref{uv} is the
following: denoting with $\bs{\hat{e}}_1$ and $\bs{\hat{e}}_2$ the
standard unit vectors of $\mathds{R}^2$, the representation
generated by the two generators $U_N=W_N(\bs{\hat{e}}_1)$ and
$V_N=W_N(\bs{\hat{e}}_2)$ and the one generated by $\Theta_N
\pt{U_N}=W_N(A\,\bs{\hat{e}}_1)$ and $\Theta_N
\pt{V_N}=W_N(A\,\bs{\hat{e}}_2)$ must be unitarily equivalent; in
other words the two representations must be labeled by the same $u$
and $v$. According to~\eqref{FoldUV}, this can be expressed by
\begin{equation}
{\pq{W_N(\bs{\hat{e}}_1)}}^N={\pq{W_N(A\bs{\hat{e}}_1)}}^N
\qquad\text{and}\qquad
{\pq{W_N(\bs{\hat{e}}_2)}}^N={\pq{W_N(A\bs{\hat{e}}_2)}}^N
\label{UVcond}\ ;
\end{equation}
the latter equation restrict the possible couples $(u,v)$ available
and leads to~\eqref{uv}.

An important set of matrices $A$, originally called ``set of
quantizable maps'' and characterized by $\pt{u,v}=\pt{0,0}$, is also
important for historical reasons, indeed it was the set used by
Berry and Hannay~\cite{Ber80:1} to develop the first quantization of
Cat Maps. Recent developments of Berry's approach to quantization
can be found in~\cite{Kea91:1,Kea00:1,Mez02:1}.

Further, relation~\eqref{Weyl3} is also preserved since the
condition \mbox{$\det\pt{A}=1$} guarantees that the symplectic form
remains invariant, i.e.\ $\sigma(A\bs{n}, A\bs{m}) =
\sigma(\bs{n},\bs{m})$. Invariance of $\sigma\pt{\cdot,\cdot}$\ ,
together with~\eqref{Weyl222-333}, also allows
equation~\eqref{dyncat} to hold true for all
$\bs{p}\,\in\mathds{Z}^2$ and not only for those in $(\mathds{Z}/ N
\mathds{Z})^2$.

\noindent Many other useful relations can be obtained by using the
explicit expression
\begin{equation}
 W_N(\bs{n})\, \vert j\rangle = \exp\pt{\frac{i\pi}{N}(-n_1n_2+ 2n_1 u+ 2n_2
 v)}\, \exp\pt{-\frac{2i\pi}{N}jn_2}\, \vert j+n_1\rangle\quad .
\label{Weyl4}
\end{equation}
In particular, from~(\ref{Weyl4}) one readily derives
 the decomposition
\begin{equation}
 \c M_N \ni X = \sum_{\bs{m}\,\in(\mathds{Z}/ N \mathds{Z})^2}
 \tau_N \Bigl(X\,W_N(-\bs{m})\Bigr)\, W_N(\bs{m})\quad,
\label{Weyl9}
\end{equation}
while from equation~(\ref{Weyl3}) one gets
\begin{equation*}
 \pq{W_N(\bs{n}), W_N(\bs{m})} = 2i\sin\pt{ \frac{\pi}{N}\,\sigma(\bs{n},\bs{m}) }\, W_N(\bs{n}+\bs{m})\quad ,
\end{equation*}
which suggests that the $\hbar$--like parameter is $1/N$ and that
the classical limit correspond to $N\to\infty$\;. In the following
section, we set up a \co{CS} technique suited to study classical cat
maps as limits of quantized cats.

\subsection{Coherent states for cat maps}
\label{cohstat}

We shall construct a \co{CS}--family $\{\;\vert C_N(\bs{x})\rangle
\mid \bs{x}\in\mathds{T}^2\;\}$ on the 2--torus by means of the
discrete Weyl group. We define
\begin{subequations}
\label{cohh}
\begin{equation}
 \vert C_N(\bs{x})\rangle \coloneqq W_N(\floor{N \bs{x}})\, \vert C_N\rangle\quad,
\label{coh1}
\end{equation}
where $\floor{N\bs{x}} = (\floor{Nx_1},\floor{Nx_2})$,
$0\le\floor{Nx_i}\le N-1$ is the largest integer smaller than $Nx_i$
and the reference vector $\vert C_N\rangle$ is chosen to be
\begin{equation}
 \vert C_N\rangle = \sum_{j=0}^{N-1} C_N(j)\vert j\rangle\qquad,\qquad
 C_N(j)\coloneqq \frac{1}{2^{(N-1)/2}}\sqrt{\binom{N-1}{j}}\quad.
\label{coh2}
\end{equation}
\end{subequations} \textsf{Measurability} and \textsf{normalization}
are immediate, \textsf{overcompleteness} comes as follows. Let $Y$
be the operator in the left hand side of Definition~\ref{coh}.3.\\If
$\tau_N(Y\, W_N(\bs{n})) = \tau_N(W_N(\bs{n}))$ for all $\bs{n} =
(n_1,n_2)$ with $0\le n_i\leqslant N-1$, then according
to~(\ref{Weyl9}) applied to $Y$ it follows that $Y=\mathds{1}$. This
is indeed the case as, using equations~\eqref{Weyl3},~\eqref{cohh}
and $N$--periodicity,
\begin{align}
 \tau_N(Y\, W_N(\bs{n}))
 &= \int_{\mathds{T}^2} \mathrm{d}\bs{x}\, \langle{}C_N(\bs{x}), W_N(\bs{n})\, C_N(\bs{x}) \rangle
\nonumber \\
 &= \int_{\mathds{T}^2} \mathrm{d}\bs{x}\, \exp{\pt{\frac{2\pi i}{N}\sigma\pt{\bs{n},\floor{N\bs{x}}}}}\, \langle{} C_N, W_N(\bs{n})\, C_N\rangle
\nonumber \\
 &= \frac{1}{N^2} \sum_{\bs{p}\,\in(\mathds{Z}/ N \mathds{Z})^2} \exp{\pt{ \frac{2\pi
 i}{N}\sigma(\bs{n},\bs{p})}}\, \langle{} C_N, W_N(\bs{n})\, C_N\rangle
\nonumber \\
 &= \tau_N (W_N(\bs{n}))\quad.
\label{coh3}
\end{align}
In the last line we used that when $\bs{x}$ runs over
$\mathds{T}^2$, $\floor{Nx_i}$,  $i=1,2$ runs over the set of
integers $0,1,\ldots, N-1$.

The proof the \textsf{localization} property in Definition~\ref{coh}
is more technical and requires several steps: the willing reader can
find it in~\cite{Ben03:1,Cap04:1}.

\section{Quantum and classical time evolutions}
\label{s5}

One of the main issues in the semi--classical analysis is to compare
if and how the quantum and classical time evolutions mimic each
other when a quantization parameter goes to zero.

In the case of classically chaotic quantum systems, the situation is
strikingly different from the case of classically integrable quantum
systems. In the former case, classical and quantum mechanics agree
on the level of coherent states only over times which scale as
$-\!\log\hbar$.

As before, let $T$ denote the evolution on the classical phase space
$\c X$ and $U_T$ the unitary single step evolution on
$\mathds{C}^N$, the so--called Floquet operator, which represent its
``quantization''. We formally state the semi--classical
correspondence of classical and quantum evolution using coherent
states:

\begin{condition}
\label{dynloc} \co{\upshape Dynamical localization:}
 There exists an $\alpha>0$ such that for all choices of
 $\varepsilon>0$ and $d_0>0$ there exists an $N_0\in\mathds{N}$ with the
 following property: if  $N > N_0$ and $k\le \alpha \log N$, then $N
 |\langle{} C_N(\bs{x}), U_T^k\,C_N(\bs{y})\rangle|^2 \le \varepsilon$ whenever $d(T^k\bs{x},\bs{y})
 \ge d_0$.
\end{condition}

\medskip
\begin{remark}
The condition of \textsf{dynamical localization} is what is expected
of a good choice of coherent states, namely, on a time scale
logarithmic in the inverse of the semi--classical parameter,
evolving \co{CS} should stay localized around the classical
trajectories. Informally, when $N\to\infty$, the quantities
\begin{equation} \label{dyn-loc} K_{k}(\bs{x},\bs{y}) \coloneqq
\langle{} C_N(\bs{x}), U_T^k C_N(\bs{y})\rangle
\end{equation}
should behave as if
$N|K_k(\bs{x},\bs{y})|^2\simeq\delta(T^k\bs{x}-\bs{y})$ (note that
this hypothesis makes our quantization consistent with the notion of
\emph{regular quantization} described in Section V
of~\cite{Slo94:1}). The constraint $k\le \alpha\log N$ is typical of
hyperbolic classical behavior and comes heuristically as follows.
The maximal localization of coherent states cannot exceed the
minimal coarse--graining dictated by $1/N$; if, while evolving,
\co{CS} stayed localized forever around the classical trajectories,
they would get more and more localized along the contracting
direction. Since for hyperbolic systems the increase of localization
is exponential with Lyapunov exponent $\log\lambda_{+}>0$, this sets
the upper bound, better known as logarithmic \emph{breaking--time},
and indicates that $\alpha\simeq1/\log\lambda_{+}$.
\end{remark}

\begin{proposition}
\label{prop3}
 Let $(\c M_N ,\Theta_N,\tau_N)$ be a general quantum
 dynamical system as defined in Section~\ref{s3} and suppose
 that it satisfies Condition~\ref{dynloc}. Let $\|X\|_2 \coloneqq \sqrt{
 \tau_N(X^*X)}$, $X\in \c M_N$ denote the normalized
 Hilbert--Schmidt norm. In the ensuing topology
 \begin{equation}
  \lim_{\substack{k,\ N\to\infty \\k< \alpha\log N}} \| \Theta_N^k
  \circ \gamma_{N \infty}(f) - \gamma_{N \infty} \circ \Theta^k(f)
  \|_2 = 0\quad.\label{added}
 \end{equation}
\end{proposition}

\smallskip

\begin{remark}
The above proposition, whose proof can be found
in~\cite{Ben03:1,Cap04:1}, can be seen as a modification of the
so--called Egorov's property (see~\cite{Mar99:1}), and gives the
strength of the non--commutativity of classical and time limits when
the classical system has a positive Lyapunov exponent. The same
(logarithmic) scaling for the \emph{breaking--time} has been found
numerically in~\cite{Ben04:1} also for discrete classical cat maps,
converging in a suitable classical limit to continuous cat maps.
Analogously, similar analysis~\cite{Ben05:1} has been performed on
sequences of discrete approximants of discontinuous automorphisms on
the 2--torus, known as Sawtooth maps, and the logarithmic
breaking--time has been recovered there too.
\end{remark}

We shall not prove the \textsf{dynamical localization}
condition~\ref{dynloc} for the quantum cat maps, but a direct
derivation of formula~\eqref{added}, based on the simple
expression~(\ref{dyncat}) of the dynamics when acting on Weyl
operators, is available in~\cite{Ben03:1,Cap04:1} and reads as
follows
\begin{proposition}
\label{prop3cat}
 Let $(\c M_N ,\Theta_N,\tau_N)$ be a sequence of quantum cat maps
 tending with $N\to\infty$ to a classical cat map with Lyapunov exponent
 $\log\lambda_{+}$; then
 \begin{equation*}
  \lim_{\substack{k,\ N\to\infty \\k< \log N/(2\log\lambda_{+})}} \|
  \Theta_N^k \circ \gamma_{N \infty}(f) - \gamma_{N \infty} \circ
  \Theta^k(f) \|_2 = 0\quad ,
  \end{equation*}
where $\|\,\cdot\,\|_2$ is the Hilbert--Schmidt norm of Proposition
5.1.
\end{proposition}

\medskip\section{Dynamical Entropies}\vspace{6mm}\label{s6}
Intuitively, one expects the instability proper to the presence of a
positive Lyapunov exponent to correspond to some degree of
unpredictability of the dynamics: classically, the metric entropy of
Kolmogorov provides the link~\cite{For92:1}.
\subsection{Kolmogorov Metric Entropy}
\label{KSME}\vspace{3mm} For continuous classical systems $({\cal
X},T,\mu)$ such as those introduced in Section~\ref{s2}, the
construction of the dynamical entropy of Kolmogorov is based on
subdividing $\cal X$ into measurable disjoint subsets
${\left\{E_\ell\mid\ell=1,2,\cdots, D\;\right\}}$ such that
$\bigcup_\ell E_\ell={\cal X}$ which form finite partitions (coarse
graining) ${\cal E}$.

Under the dynamical maps $T:{\cal X}\to{\cal X}$ , any given ${\cal
E}$ evolves into $T^{j}({\cal E})$ with atoms $\displaystyle
T^{-j}(E_\ell)=\{\bs{x}\in{\cal X} \mid  T^j\bs{x}\in E_\ell\}$; one
can then form finer partitions
\begin{alignat}{6}
{\cal E}_{[0,n-1]}& \coloneqq\bigvee_{j=0}^{n-1}T^{j}({\cal E})&\  &
=\
 {\cal E}&&\bigvee \ \ \ T({\cal E})&&\bigvee&&\cdots&&\bigvee
 \ T^{n-1}({\cal E})\notag
\intertext{whose atoms}
 E_{i_0\,i_1\cdots i_{n-1}}&\coloneqq
 \bigcap_{j=0}^{n-1}T^{-j}E_{i_j}&&=\
 E_{i_0}&&\bigcap T^{-1}(E_{i_1})&&\bigcap &&\cdots&&\bigcap
 T^{-n+1}(E_{i_{n-1}})\notag
\intertext{have volumes} \mu_{i_0\,i_1\cdots
i_{n-1}}&\coloneqq\mu\left( E_{i_0\,i_1\cdots i_{n-1}}
\right)&&\cdot&&&&&&&& \label{KSE_1}
\end{alignat}\\[-5.5ex]
\begin{definition}{}\label{stringhe}
We shall set $\bs{i}=\pg{i_0\,i_1\cdots i_{n-1}}$ and denote by
$\Omega_D^n$ the set of $D^n$ n\_tuples with $i_j$ taking values in
$\pg{1, 2, \cdots, D}$.
\end{definition}
\noindent The atoms of the partitions ${\cal E}_{[0,n-1]}$ describe
segments of trajectories up to time $n$ encoded by the atoms of
${\cal E}$ that are traversed at successive times; the volumes
$\mu_{\bs{i}}=\mu\pt{E_{\bs{i}}}$ corresponds to probabilities for
the system to belong to the atoms
$E_{i_0},E_{i_1},\cdots,E_{i_{n-1}}$ at successive times $0\leqslant
j\leqslant n-1$. The $n$\_tuples $\bs{i}$ by themselves provide a
description of the system in a symbolic dynamic.

The richness in diverse trajectories, that is the degree of
irregularity of the motion (as seen with the accuracy of the given
coarse-graining) correspond intuitively to our idea of
``complexity'' and can be better measured by the Shannon
entropy~\cite{Ale81:1}
\begin{equation}
S_\mu({\cal
E}_{[0,n-1]})\coloneqq-\sum_{\bs{i}\in\Omega_D^n}\mu_{\bs{i}}
\log\mu_{\bs{i}}\ . \label{KSE_2}
\end{equation}
In the long run, ${\cal E}$ attributes to the dynamics an entropy
per unit time--step
\begin{equation}
h_\mu(T,{\cal E})\coloneqq\lim_{n\to\infty}\frac{1}{n}S_\mu({\cal
E}_{[0,n-1]})\ . \label{KSE_3}
\end{equation}
This limit is well defined~\cite{Kat99:1} and the ``average entropy
production'' $h_\mu(T,{\cal E})$ measure how predictable the
dynamics is on the coarse grained scale provided by the finite
partition ${\cal E}$. To remove the dependence on ${\cal E}$, the
Kolmogorov--Sinai entropy $h^{\text{\upshape \co{KS}}}_\mu(T)$ of
$({\cal X},T,\mu)$ (or \co{KS}--entropy) is defined as the supremum
over all finite measurable partitions~\cite{Kat99:1,Ale81:1}:
\begin{equation}
h^{\text{\upshape \co{KS}}}_\mu(T)\coloneqq\sup_{{\cal
E}}h_\mu(T,{\cal E})\qquad\ \cdot \label{KSE_4}
\end{equation}
For the automorphisms of the 2-torus, we have the well-known
result~\cite{Kat99:1}:

\begin{proposition}
\label{prop4.1}
 Let $(\g M_\mu,\Theta,\omega_\mu)$ be as in Definition~\ref{ccat-map},
 then $h^{\text{\upshape \co{KS}}}_\mu(T) = \log\lambda_+$.
\end{proposition}

\subsection{Quantum Dynamical Entropies}\label{QDE}

\noindent The idea behind the notion of dynamical entropy is that
information can be obtained by repeatedly observing a system in the
course of its time evolution. Due to the uncertainty principle, or,
in other words, to non-commutativity, if observations are intended
to gather information about the intrinsic dynamical properties of
quantum systems, then  non-commutative extensions of the
\co{KS}-entropy ought first to decide whether quantum disturbances
produced by observations have to be taken into account or not.

Concretely, let us consider a quantum system described by a density
matrix $\rho$ acting on a Hilbert space $\g H$.  Via the wave packet
reduction postulate, generic measurement processes may be described
by finite sets $\c Y = \{y_1, y_2,\ldots, y_D\}$ of bounded
operators $y_j\in \c B(\g H)$ such that $\sum_j y_j^*
y_j^{\phantom{*}} = \mathds{1}$. These sets are called
\co{partitions of unity} ({\textsf{p.u.}}, for sake of shortness)
and describe the change in the state of the system caused by the
corresponding measurement process:
\begin{equation}
\label{18}
 \rho \mapsto \Gamma^*_{\c Y}(\rho) := \sum_j y_j\, \rho\, y^*_j.
\end{equation}
It looks rather natural to rely on partitions of unity to describe
the process of collecting information through repeated observations
of an evolving  quantum system~\cite{Ali94:1}. Yet, most of these
measurements interfere with the quantum evolution, possibly acting
as a source of unwanted extrinsic randomness. Nevertheless, the
effect is typically quantal and rarely avoidable.  Quite
interestingly, as we shall see later, pursuing these ideas leads to
quantum stochastic processes with a quantum dynamical entropy of
their own, the \co{ALF}-entropy, that is also useful in a classical
context.

An alternative approach~\cite{Con87:1} leads to the dynamical
entropy of Connes, Narnhofer and
Thirring~\cite{Con87:1}(\co{CNT}--entropy). This approach lacks the
operational appeal of the \co{ALF}-construction, but is intimately
connected with the intrinsic relaxation properties of quantum
systems~\cite{Con87:1,Nar92:1} and possibly useful in the rapidly
growing field of quantum communication. The \co{CNT}-entropy is
based on decomposing quantum states rather than on reducing them as
in~(\ref{18}).  Explicitly, if the state $\rho$ is not a one
dimensional projection, any partition of unity $\c Y$ yields a
decomposition
\begin{equation}
\label{19}
 \rho = \sum_j \tr \bigl(\rho\, y^*_jy_j^{\phantom{*}}\bigr)\, \frac{\sqrt\rho\,
 y^*_jy_j\sqrt\rho} {\tr\bigl(\rho\, y^*_jy_j\bigr)}\quad\cdot
\end{equation}
When $\Gamma^*_{\c Y}(\rho) = \rho$, reductions also provide
decompositions, but not in general.

A different kind of wave packet reduction is the starting point for
constructing the \emph{coherent states
entropy}~\cite{Slo94:1,Slo98:1} (in the following \co{\upshape
CS}--entropy, for short), in fact based on coherent states
$\coh{N}{\bs{x}}$ as the ones introduced Definition~\ref{coh}.

The map
\begin{equation}
{\c I}\pt{E}\pt{\rho}\coloneqq N
\int_{E}\coh{N}{\bs{x}}\lcoh{N}{\bs{x}}\ \rho\
\coh{N}{\bs{x}}\lcoh{N}{\bs{x}}\;\mu\pt{\mathrm{d} \bs{x}}\quad
\label{inst},
\end{equation}
for a measurable subset $E\subset{\c X}$ and an operator $\rho$, is
called an \emph{instrument}: it describe the change in the state
$\rho$ of the system caused by an $E$--dependent measurement process
(compare with~\eqref{18}), actually a double approximate measurement
in the phase space. Repeated measurement, taken stroboscopically
during the dynamical evolution and performed with different
instrument ${\c I}(E_{i_j})$ labeled by different elements $E_{i_j}$
of a partition $\c E$, map the input state $\rho$ into many possible
output $\{\;\rho_{\bs{i}}\mid\bs{i}\in\Omega_D^n\;\}$, which in turn
can be mapped into many positive numbers
$\{\;\mathds{R}^+\ni\omega_{\bs{i}}\coloneqq\omega
\pt{\rho_{\bs{i}}}\mid\bs{i}\in\Omega_D^n\;\}$ summing up to one.
Now we have once more the correspondence between strings
$\bs{i}\in\Omega_D^n$ and probability $\omega_{\bs{i}}$, in other
word we end up with a probability space and a similar reasoning
leading us in Section~\ref{KSME} to the \co{KS} invariant, can now
be used for constructing the \co{CS}--entropy.

\subsection{\co{\upshape ALF}--entropy}

The idea underlying the \co{ALF}--entropy is that the evolution of a
quantum dynamical system can be modeled by repeated  measurements at
successive equally spaced times, the measurements corresponding to
\textsf{p.u.} as in equation~\eqref{18}.

Such a construction associates a quantum dynamical system  with a
symbolic dynamics corresponding  to the right--shift along a quantum
spin half--chain~\cite{Tuy97:1}.

Generic {\co{p.u.}} $\c Y = \{y_1, y_2,\ldots, y_D\}$ need not
preserve the state, but disturbances are kept under control by
suitably selecting the subalgebra of observables ${\g M}_0\ni
y_j$\;. The construction of the \co{ALF}--entropy for a quantum
dynamical system $(\g M,\Theta,\omega)$ can be resumed as follows:

\begin{itemize}
\item
 One selects a $\Theta$--invariant subalgebra $\g M_0 \subseteq \g M$
  and a
 {\co{\upshape p.u.}} $\c Y = \{ y_1,\ldots, y_D \}$ of finite
 size $D$ with $y_j\in\g M_0$. After $j$ time steps $\c Y$ will have
 evolved into another {\co{\upshape p.u.}}
 from $\g M_0$: $\Theta^j(\c Y) \coloneqq \{ \Theta^j(y_1), \Theta^j(y_2),\ldots,
 \Theta^j(y_D) \} \subset \g M_0$.
\item
 Every {\co{\upshape p.u.}} $\c Y$ of size $D$ gives rise to an
 $D$--dimensional density matrix
 \begin{equation}
  \rho[\c Y]_{i,j} \coloneqq \omega(y^*_jy_i^{\phantom{*}}),
 \label{29}
 \end{equation}
 with von~Neumann entropy $H_\omega[\c Y] \coloneqq S(\rho[\c Y])=-\tr\Big(\rho[{\cal Y}]\log\rho[{\cal Y}]\Big)$\;.
\item
 Given two partitions of unit
${\cal Y}=\{y_1,y_2,\ldots,y_D\}$, ${\cal
Z}=\{z_1,z_2,\ldots,z_B\}$, of size $D$, respectively $B$, one gets
a finer partition of unit of size $BD$ as the set
\begin{equation}
{\cal Y}\circ {\cal Z} \coloneqq\{ \;y_1 z_1,\ldots,y_1 z_B\,; y_2
z_1,\ldots,y_2 z_B\,;\ldots; y_D z_1,\ldots,y_D z_B \;\}\quad\cdot
\label{AFE_4}
\end{equation}
\item
 Given a size~$D$ {\co{\upshape p.u.}} $\c Y$ and the \co{\upshape ordered
 time refinements}
 \begin{equation}
 \c Y^{[0,n-1]} \coloneqq \Theta^{n-1}(\c Y) \circ \Theta^{n-2}(\c Y)
 \circ\cdots\circ \c Y,
 \label{31}
 \end{equation}
 the $D^n\times D^n$ density matrices $\rho^{[0,n-1]}_{\c Y} \coloneqq \rho\pq{\c Y^{[0,n-1]}}$ define
 states on the $n$--fold tensor product $\c M_D^{\otimes n}$ of
 $D$--dimensional matrix algebras $\c M_D$.
\item
 Given a {\co{p.u.}} $\c Y$ of size $D$, let
 $\Phi_{\c Y}: \c M_D\otimes\g M \mapsto \g M$ and $e_M: \g
 M\mapsto\g M$, with $M\in\c M_D$, be linear maps defined by
 \begin{equation}
  \Phi_{\c Y}(M\otimes x) \coloneqq \sum_{i,j} y_i^*x\,y_j\, M_{ij}
  \quad\text{and}\quad
  e_M(x) \coloneqq \sum_{i,j} y^*_i\Theta(x)\,y_j\, M_{ij}\quad.
 \label{32}
 \end{equation}
 $\Phi_{\c Y}$ is a completely positive unital map, while $e_\mathds{1}(\mathds{1})=\mathds{1}$.
 One readily computes
 \begin{equation*}
  \omega\Bigl( e_{M_0} \circ e_{M_1}\cdots \circ
  e_{M_{n-1}}(\mathds{1}) \Bigr) = \tr\Bigl( \rho^{[0,n-1]}_{\c Y}\, M_0\otimes
  M_1\cdots\otimes M_{n-1} \Bigr).
 \end{equation*}
\end{itemize}

The states $\rho_{\cal Y}^{[0,n-1]}$ are compatible:
 \begin{equation*}
\rho^{[0,n-1]}_{\cal Y}\text{\large$\upharpoonright$}\;{\cal
M}_D^{[0,n-2]}= \rho^{[0,n-2]}_{\cal Y}\quad,\quad \text{where}\quad
{\cal M}_D^{[0,n-2]}\coloneqq\bigotimes_{\ell=0}^{n-2}({\cal
M}_D)_\ell \quad,
 \end{equation*}
and define a global state $\rho_{\cal Y}$ on the quantum spin chain
$\displaystyle {\cal
M}_D^\infty\coloneqq\otimes_{\ell=0}^\infty({\cal M}_D)_\ell$.

Thus it is possible to associate with the quantum dynamical system
$(\g M,\Theta,\omega)$ a symbolic dynamics which amounts to the
right--shift
 $\displaystyle \sigma:{({\cal M}_D)}_\ell\mapsto{({\cal
M}_D)}_{\ell+1}$ along the quantum spin half--chain.\\
Non-commutativity becomes evident when we check whether $\rho_{\cal
Y}$ is shift-invariant. This requires $\omega\Big(\sum_\ell y_\ell^*
x\: y_\ell^{\phantom{*}}\Big)=\omega(x)$ for all $x\in{\cal M}$.
Note that this is the case in which $\rho \mapsto \Gamma^*_{\c
Y}(\rho) =\rho$ \ (see. equation~\eqref{18}).
\begin{definition}
 The \co{\upshape ALF}--entropy of a quantum dynamical system $(\g M,\Theta,\omega)$
 is
\begin{subequations}
\label{AFE_7}
\begin{align}
h^{\co{ALF}}_{\omega,{\cal M}_0}(\Theta) & \coloneqq\sup_{{\cal
Y}\subset{\cal M}_0} h^{\co{ALF}}_\omega(\Theta,{\cal Y})\ ,
\label{AFE_7a} \\
\text{where}\qquad \qquad h^{\co{ALF}}_\omega(\Theta,{\cal Y}) &
\coloneqq\limsup_n \frac{1}{n} H_{\omega}\Big[{\cal
Y}^{[0,n-1]}\Big]\ \cdot \label{AFE_7b}
\end{align}
\end{subequations}
\end{definition}
Like the metric entropy of a partition ${\cal E}$, also the
\co{ALF}--entropy of a partition of unit ${\cal Y}$ can be
physically interpreted as an asymptotic \emph{entropy production}
relative to a specific coarse--graining.
\subsection{Comparison of dynamical entropies}
In this section we outline some of the main common features of many
different dynamical entropies, taking the \co{ALF} as a reference
example, because of its conceptual simplicity. Here, we just sketch
such a features, emphasizing those parts that are important to the
study of the classical limit of quantum dynamical entropies
(\co{QDE}).

The first thing to notice is that the any \co{QDE} must coincide
with the \co{KS}--entropy when $\g M =\g M_\mu$ is the Abelian
von~Neumann algebra $\g L^\infty_\mu(\c X)$ and $(\g
M,\Theta,\omega)$ represents a classical dynamical system.

The next observation is that when, as for the quantized hyperbolic
automorphisms of the torus considered in this paper, $\g M$ is a
finite--dimensional algebra, both the \co{CNT}-- and the
\co{ALF}--entropy are zero, see~\cite{Con87:1,Ali94:1}.
Consequently, if we decide to take the strict positivity of
\co{ALF}-- or \co{CNT}--entropy as a signature of quantum chaos,
quantized hyperbolic automorphisms of the torus cannot be called
chaotic.

However, the latter observation is not as general as the former.
There exist many alternative definitions (different from \co{ALF}
and \co{CNT}), and some of them need no to be equal to zero for all
quantum systems defined on a finite dimensional Hilbert space: an
interesting example is represented by the \co{\upshape CS}--entropy
introduced in~\cite{Slo94:1}.

From the previous considerations, it is clear that the main field of
application of the \co{CNT}-- and \co{ALF}--entropies are infinite
quantum systems, where the  differences between the two come to the
fore~\cite{Ali95:1}.

The complete proofs of the above facts can be found
in~\cite{Con87:1} for the \co{CNT}--, in~\cite{Ali96:1,Ali94:1} for
the \co{ALF}-- and in~\cite{Slo94:1} for the \co{CS}--entropy. Here
we just state more rigorously the above observations, in the case of
the \co{ALF}--entropy, in the two subsequent
Proposition~\ref{alf_clas_1} and Proposition~\ref{prop4.3}.

\begin{proposition}\label{alf_clas_1}
 Let $(\g M_\mu,\Theta,\omega_\mu)$ represent a classical dynamical
 system. Then, with the notations of the previous sections
 \begin{equation*}
   h^{\text{\upshape \co{ALF}}}_{(\omega_\mu, \g M_\mu)}(\Theta) = h^{\text{\upshape \co{KS}}}_\mu(T)\quad.
 \end{equation*}
\end{proposition}

In the particular case of the hyperbolic automorphisms of the torus,
we may restrict our attention to {\co{p.u.}} whose elements belong
to the $\ast$--algebra $\c D_\mu$ of complex functions $f$ on
$\mathds{T}^2$ such that the support of $\hat f$ is bounded:
\begin{equation*}
 h^{\text{\upshape \co{KS}}}_\mu(T) = h^{\text{\upshape \co{ALF}}}_{(\omega_\mu, \g M_\mu)}(\Theta) =
 h^{\text{\upshape \co{ALF}}}_{(\omega_\mu, \c D_\mu)}(\Theta).
\end{equation*}
Remarkably, the computation of the classical \co{KS}--entropy via
the quantum mechanical \co{ALF}--entropy yields a proof of
Proposition~\ref{prop4.1} that is much simpler than the standard
ones~\cite{Arn68:1,Wal82:1}.

\begin{proposition}
\label{prop4.3}
 Let $(\g M,\Theta,\omega)$ be a quantum dynamical system with
 $\g M$, a finite dimensional C*--algebra, then,
 \begin{equation*}
  h^{\text{\upshape \co{ALF}}}_{(\omega,\g M)}(\Theta) = 0\quad.
 \end{equation*}
\end{proposition}

\section{Classical limit of quantum dynamical entropies}
\label{s7}

Proposition~\ref{prop4.3} confirms the intuition that finite
dimensional, discrete time, quantum dynamical systems, however
complicated the distribution  of their quasi--en\-ergies might be,
cannot produce enough information  over large times to generate a
non--vanishing entropy per unit time. This is due to the fact that,
despite the presence of almost random features over finite
intervals, the time evolution cannot bear random signatures if
watched long enough, because almost periodicity would always prevail
asymptotically.

However, this does not mean that the dynamics may not be able to
show a significant entropy rate over finite interval of times, these
being typical of the underlying dynamics. Here we show that
underlying classical chaos plus Hilbert space finiteness make a
characteristic logarithmic time scale emerge over which these
systems can be called chaotic. This is precisely the content of the
next Theorem~\ref{maint}, whose proof can be found
in~\cite{Ben03:1,Cap04:1}.

\begin{theorem}\label{maint}
 Let $(\c X,T,\mu)$ be a classical dynamical system
 which is the classical limit of a sequence of finite dimensional
 quantum dynamical systems $(\c M_N,\Theta_N,\tau_N)$.
 We also assume that the dynamical localization
 condition~\ref{dynloc} holds. If
 \begin{enumerate}
 \item
  $\c E = \{ E_1, E_2,\ldots, E_{D-1} \}$ is a finite measurable
  partition of $\c X$,
 \item
  $\c Y_N = \{ y_1, y_2,\ldots, y_D \}$ is a bistochastic partition of
  unity, which is the quantization  of the previous partition, namely
  $y_i = \gamma_{N\infty}(\chi_{E_i})$  for $i=1,2,\ldots,D-1$  and
  $y_D \coloneqq \sqrt{\mathds{1}-\sum_{i=0}^{D-1} y_i^*y_i^{\phantom{*}}}$,
 \end{enumerate}
 then there exists an $\alpha$ such that
 \begin{equation*}
  \lim_{\substack{k,N\to\infty \\ k\le\alpha\log N}} \frac{1}{k}
  \left| H[\c Y_N^{(k)}] - S_\mu(\c E^{(k)}) \right| = 0\quad.
 \end{equation*}
\end{theorem}
A similar phenomenon has been proved both for the
\co{CNT}--entropy~\cite{Ben03:1,Cap04:1} and for the
\co{CS}--entropy~\cite{Cap05:1}, although in this case a different
kind of dynamical system has been studied. Nevertheless, the proof
of convergence of \co{CS}--entropy to the \co{KS} invariant only
makes use of dynamical localization condition~\ref{dynloc} so that,
after an appropriate substitution of similar terms,
Theorem~\ref{maint} can be extended to both \co{CNT}-- and \co{CS}--
entropies.

The dynamical localization condition~\ref{dynloc} has been
extensively used in all the proofs mentioned in this Section, and
the results here presented strongly do depend on it. Once the
framework in which sequences of quantum approximants approach their
classical limit has been settled, by an appropriate Egorov
convergence, like the one in Propositions~\ref{prop3}
and~\ref{prop3cat}, we still let room for bizarre behaviors in the
entropy production. Condition~\ref{dynloc} remove such a freedom and
extend the convergence from observables to dynamical entropies.

\section{Conclusions and outlook}
We have reviewed how quantum dynamical entropies reproduce the
Kolmo\-gorov--Sinai invariant in quantum systems too, provided that
we observe a strongly chaotic system on a very short logarithmic
time scale. However, due to the discreteness of the spectrum of the
quantizations, we know that saturation phenomena will appear. It
would be interesting to study the scaling behavior of the quantum
dynamical entropies in the intermediate region between the
logarithmic breaking time and the Heisenberg time. This will,
however, require quite different techniques than the coherent states
approach, indeed the Ehrenfest time, whose scaling is the same as
the breaking time here described, set the upper bound of
semi--classical technology.

\section*{Acknowledgements}
{   It is a pleasure to thank F. Benatti and K. \.Zyczkowski for
stimulating discussions. The author  acknowledges financial support provided by the
 \textsf{EU} Marie Curie Host Fellowships for Transfer of Knowledge
 Project \co{COCOS} (contract number \textsf{MTKD}--\textsf{CT}--2004--517186)
 and the \co{SFB}/Transregio--12 project financed by \co{DFG}.}


\begin{thebibliography}{99}
\addcontentsline{toc}{section}{References}

\bibitem{Dev89:1}
R.~Devaney,
\newblock {\it An Introduction to Chaotic Dynamical Systems},
\newblock Addison--Wesley, Reading, MA, 1989.

\bibitem{Wig90:1}
S.~Wiggins,
\newblock {\it Dynamical Systems and Chaos},
\newblock Springer--Verlag, New York, 1990.

\bibitem{Kat99:1}
A.~Katok and B.~Hasselblatt,
\newblock {\it Introduction to the Modern Theory of Dynamical Systems},
\newblock {\it Encyclopedia of Mathematics and its Applications}, Cambridge
  University Press, Cambridge, 1999.

\bibitem{Sch95:1}
H.G. Schuster,
\newblock {\it Deterministic Chaos},
\newblock VCH, Weinheim, 3rd edition, 1995.

\bibitem{Gia89:1}
M.{--J}. Giannoni, A.~Voros and J.~{Zinn--Justin}, editors,
\newblock {\it Chaos and Quantum Physics}, volume 1989, Les Houches Session LII
  of {\it Les Houches Summer School of Theoretical Physics}, Amsterdam, 1991,
  North--Holland.

\bibitem{Cas95:1}
G.~Casati and B.~Chirikov,
\newblock {\it Quantum Chaos. Between Order and Disorder},
\newblock Cambridge University Press, Cambridge, 1995.

\bibitem{Zas85:1}
G.M. Zaslawsky,
\newblock {\it Chaos in Dynamic Systems},
\newblock Harwood Academic Publ., Chur, 1985.

\bibitem{For92:1}
J.~Ford and M.~Ilg,
\newblock Eigenfunctions, eigenvalues, and time evolution of finite, bounded,
  undriven, quantum systems are not chaotic,
\newblock {\it Phys. Rev. A} {\bf 45}{\bf (9)}, 6165--6173 (1992).

\bibitem{Zas91:1}
G.M. Zaslawsky,
\newblock From classical to quantum chaos,
\newblock in {\it Quantum Chaos}, edited by H.~A. Cerdeira, R.~Ramaswamy, M.~C.
  Gutzwiller and G.~Casati, page~32, Singapore, 1991, World Scientific.

\bibitem{For91:1}
J.~Ford, G.~Mantica and G.H. Ristow,
\newblock The {Arno}ld cat-failure of the correspondence principle,
\newblock {\it Physica D} {\bf 50}, 493--520 (1991).

\bibitem{For92:2}
J.~Ford and G.~Mantica,
\newblock Does quantum mechanics obey the correspondence principle? --- is it
  complete?,
\newblock {\it Am.\ J.\ Phys.} {\bf 60}, 1086--1098 (1992).

\bibitem{Arn68:1}
V.I. Arnold and A.~Avez,
\newblock {\it Ergodic Problems of Classical Mechanics},
\newblock Benjamin, New York, NY, 1968.

\bibitem{Wal82:1}
P.~Walters,
\newblock {\it An Introduction to Ergodic Theory}, volume~79 of {\it Graduate
  Text in Mathematics},
\newblock Springer--Verlag, Berlin, 1982.

\bibitem{Ber79:1}
M.V. Berry, N.L. Balazs, M.~Tabor and A.~Voros,
\newblock Quantum maps,
\newblock {\it Ann. of Phys.} {\bf 122}, 26--63 (1979).

\bibitem{Deg93:1}
M.~Degli Esposti,
\newblock Quantization of the orientation preserving automorphisms of the
  torus,
\newblock {\it Ann.\ Inst.\ Henri Poincar{\'e}} {\bf 58}, 323--34 (1993).

\bibitem{Man87:1}
R.~Ma{\~n}{\'e},
\newblock {\it Ergodic Theory and Differentiable Dynamics},
\newblock Springer--Verlag, Berlin, 1987.

\bibitem{Con87:1}
A.~Connes, H.~Narnhofer and W.~Thirring,
\newblock Dynamic entropy of {C*-algebras} and {Von~Neumann} algebras,
\newblock {\it Comm. Math. Phys.} {\bf 112}, 691--719 (1987).

\bibitem{Ali94:1}
R.~Alicki and M.~Fannes,
\newblock Defining quantum dynamical entropy,
\newblock {\it Lett. Math. Phys.} {\bf 32}, 75--82 (1994).

\bibitem{Voi92:1}
D.~Voiculescu,
\newblock Dynamical approximation entropies and topological entropy in operator
  algebras,
\newblock {\it Comm. Math. Phys.} {\bf 144}, 443--490 (1992).

\bibitem{Acc97:1}
L.~Accardi, M.~Ohya and N.~Watanabe,
\newblock Dynamical entropy through quantum {Markov} chains,
\newblock {\it Open Sys.\ \& Information Dyn.} {\bf 4}, 71--87 (1997).

\bibitem{Slo94:1}
W.~S{\l}omczy{\'n}ski and K.~{\.Zyczkowski},
\newblock Quantum chaos: an entropy approach,
\newblock {\it J. Math. Phys.} {\bf 35}{\bf (11)}, 5674--5700 (1994).

\bibitem{Ali96:2}
R.~Alicki, D.~Makowiec and W.~Miklaszewski,
\newblock Quantum chaos in terms of entropy for a periodically kicked top,
\newblock {\it Phys. Rev. Lett.} {\bf 77}{\bf (5)}, 838--841 (1996).

\bibitem{Pog07:1}
M.~Pogorzelska and R.~Alicki,
\newblock Linear dynamical entropy and free-independence for quantized linear
  maps on the torus,
\newblock {\it J. Phys. A: Math. Theor.} {\bf 40}{\bf (13)}, 3379--3388 (2007).

\bibitem{Ali01:1}
R.~Alicki and M.~Fannes,
\newblock {\it Quantum Dynamical Systems},
\newblock Oxford University Press, Oxford, 2001.

\bibitem{Wer95:1}
R.F. Werner,
\newblock The classical limit of quantum theory,
\newblock {\it Preprint }{\tt quant-ph/9504016}, 1995.

\bibitem{Rud87:1}
W.~Rudin,
\newblock {\it Real and Complex Analysis},
\newblock McGraw--Hill, New York, 3rd edition, 1987.

\bibitem{Ben03:1}
F.~Benatti, V.~Cappellini, M.{~De~Cock}, M.~Fannes and
D.~Vanpeteghem,
\newblock Classical limit of quantum dynamical entropies,
\newblock {\it Rev. Math. Phys.} {\bf 15}{\bf (8)}, 847--875 (2003).

\bibitem{Cap04:1}
V.~Cappellini,
\newblock {\it Quantum Dynamical Entropies and Complexity in Dynamical
  Systems},
\newblock PhD thesis, University of Trieste, 2004,
\newblock {{\tt math-ph/0403035}}.

\bibitem{Per87:1}
I.~C. Percival and F.~Vivaldi,
\newblock A linear code for the sawtooth and cat maps,
\newblock {\it Physica D} {\bf 27}, 373 (1987).

\bibitem{Ben91:2}
F.~Benatti, H.~Narnhofer and G.L. Sewell,
\newblock A non-commutative version of the {Arno}ld cat map,
\newblock {\it Lett. Math. Phys.} {\bf 21}, 157--172 (1991).

\bibitem{Deb98:1}
S.~{De Bi{\`e}vre},
\newblock Chaos, quantization and the classical limit on the torus,
\newblock in {\it Proceedings of the XIVth Workshop on Geometrical Methods in
  Physics, Bialowieza (Poland) 1995}, edited by A.~Strassburger, S.~Twareque
  Ali, J.-P. Antoine, J.-P. Gazeau and A.~Odzijewicz, Warsaw, 1998, Polish
  Scientific Publisher PWN,
\newblock {\it Preprint }{\tt mp\_arc 96-191}.

\bibitem{Ber80:1}
M.~V. Berry and J.~H. Hannay,
\newblock Quantization of linear maps on a {torus -- Fresnel} diffraction by a
  periodic grating,
\newblock {\it Physica D} {\bf 1}, 267--290 (1980).

\bibitem{Kea91:1}
J.~P. Keating,
\newblock The cat maps: quantum mechanics and classical motion,
\newblock {\it Nonlinearity} {\bf 4}{\bf (2)}, 309--341 (1991).

\bibitem{Kea00:1}
J.~P. Keating and F.~Mezzadri,
\newblock Pseudo--symmetries of anosov maps and spectral statistics,
\newblock {\it Nonlinearity} {\bf 13}{\bf (3)}, 747--775 (2000).

\bibitem{Mez02:1}
F.~Mezzadri,
\newblock On the multiplicativity of quantum cat maps,
\newblock {\it Nonlinearity} {\bf 15}{\bf (3)}, 905--922 (2002).

\bibitem{Mar99:1}
J.~Marklof and Z.~Rudnick,
\newblock Quantum unique ergodicity for parabolic maps,
\newblock {\it GAFA, Geom. Funct. Anal.} {\bf 10}, 1554--1578 (2000).

\bibitem{Ben04:1}
F.~Benatti, V.~Cappellini and F.~Zertuche,
\newblock Quantum dynamical entropies in discrete classical chaos,
\newblock {\it J. Phys. A: Math. Gen.} {\bf 37}{\bf (1)}, 105--130 (2004).

\bibitem{Ben05:1}
F.~Benatti and V.~Cappellini,
\newblock Continuous limit of discrete {Sawtooth Maps} and its algebraic
  framework,
\newblock {\it J. Math. Phys.} {\bf 46}{\bf (6)}, 062702 (2005).

\bibitem{Ale81:1}
V.~M. Alekseev and M.~V. Yakobson,
\newblock Symbolic dynamics and hyperbolic dynamical systems,
\newblock {\it Phys. Rep.} {\bf 75}, 287--325 (1981).

\bibitem{Nar92:1}
H.~Narnhofer,
\newblock Quantized {Arno}ld cat maps can be entropic {K-systems},
\newblock {\it J. Math. Phys.} {\bf 33}, 1502--1510 (1992).

\bibitem{Slo98:1}
W.~S{\l}omczy{\'n}ski and K.~{\.Zyczkowski},
\newblock Mean dynamical entropy of quantum maps on the sphere diverges in the
  semiclassical limit,
\newblock {\it Phys. Rev. Lett.} {\bf 80}, 1880--1883 (1998).

\bibitem{Tuy97:1}
P.~Tuyls,
\newblock {\it Towards Quantum Dynamical Entropy},
\newblock PhD thesis, K.U.--Leuven, 1997.

\bibitem{Ali95:1}
R.~Alicki and H.~Narnhofer,
\newblock Comparison of dynamical entropies for the noncommutative shifts,
\newblock {\it Lett. Math. Phys.} {\bf 33}, 241--247 (1995).

\bibitem{Ali96:1}
R.~Alicki, J.~Andries, M.~Fannes and P.~Tuyls,
\newblock An algebraic approach to the {Kolmogorov--Sinai} entropy,
\newblock {\it Rev. Math. Phys.} {\bf 8}, 167--184 (1996).

\bibitem{Cap05:1}
V.~Cappellini,
\newblock Quantum dynamical entropies for discrete classical systems: a
  comparison,
\newblock {\it J. Phys. A: Math. Gen.} {\bf 38}{\bf (31)}, 6893--6915 (2005).
\end{thebibliography}
\end{document}